\documentclass[12pt,preprint]{aastex}
\usepackage{emulateapj5}
\usepackage{xspace}
\usepackage{graphics,graphicx}
\usepackage{array}
\usepackage{multirow}
\newcommand{\as}{\mbox{\ensuremath{.\!\!^{\prime\prime}}}}
\newcommand{\asn}{$^{\prime\prime}$\xspace}
\newcommand{\am}{$^{\prime}$\xspace}
\newcommand{\nH}{$N_{\rm H}$\xspace}
\newcommand{\etaC}{$\eta$ Car\xspace}
\newcommand{\PL}{$\Gamma$\xspace}
\newcommand{\Msun}{$M_{\odot}$\xspace}
\newcommand{\Cdof}{$C/\mathit{d.o.f.}$\xspace}

\slugcomment{}
\shorttitle{SN 2010da: A Supergiant X-ray Binary in NGC~300}
\shortauthors{Binder et al.}

\begin{document}

\title{{\it Chandra} Detection of SN 2010da Four Months After Outburst: Evidence for a Supergiant X-ray Binary in NGC~300}
\author{B. Binder\altaffilmark{1}, 
B. F. Williams\altaffilmark{1}, 
A. K. H. Kong\altaffilmark{2},
T. J. Gaetz\altaffilmark{3},  
P. P. Plucinsky\altaffilmark{3},
J. J. Dalcanton\altaffilmark{1}, 
D. R. Weisz\altaffilmark{1}
}
\altaffiltext{1}{University of Washington, Department of Astronomy, Box 351580, Seattle, WA 98195}
\altaffiltext{2}{Institute of Astronomy and Department of Physics, National Tsing Hua University, Hsinchu 30013, Taiwan}
\altaffiltext{3}{Harvard-Smithsonian Center for Astrophysics, 60 Garden Street Cambridge, MA 02138, USA}

\begin{abstract}
We present the results of a 63 ks {\it Chandra} observation of the ``supernova impostor'' SN 2010da four months after it was first observed on 25 May 2010. We detect an X-ray source at $\sim$7$\sigma$ confidence coincident with the optical location of the SN 2010da outburst. Our imaging analysis has revealed a hard central point source, surrounded by soft diffuse emission extending as far as $\sim$8\asn north of the central source. The diffuse emission has a hardness ratio, 0.35-2 keV luminosity ($\sim6\times10^{35}$ erg s$^{-1}$), and size ($\sim20$ pc) consistent with that of a supernova remnant, although the low number of counts prohibits detailed spectral modeling. Our best-fit spectral model for the hard central source is a black body ($kT = 1.79^{+0.66}_{-0.43}$ keV) with no evidence for intrinsic absorption beyond the Galactic column. We estimate the 0.3-10 keV luminosity to be 1.7$^{+0.2}_{-0.5}\times10^{37}$ erg s$^{-1}$, a factor of $\sim$25 decrease since the initial outburst four months previously. The high X-ray luminosity and slow fading rate is not consistent with this object being a single massive star undergoing an outburst; instead, we favor the scenario where the massive star powering the SN 2010da optical transient is part of a wind-fed supergiant X-ray binary system with a compact companion powering the observed X-ray emission.
\end{abstract}

\section{Introduction}
Supernova (SN) 2010da was first detected as an optical transient on 24 May 2010 in NGC~300 (Khan et al. 2010, ATel \#2632); the absolute magnitudes and spectrum (Elias-Rosa et al. 2010, ATEL \#2636) were subsequently found to be consistent with a luminous blue variable (LBV) outburst origin rather than a true SN explosion. Within hours of the initial optical detection, SN 2010da was observed by the {\it Swift} X-ray Telescope (XRT; Immler et al. 2010, ATel \#2639). A bright X-ray point source with a 0.2-10 keV luminosity of (4.5$^{+0.9}_{-2.1})\times10^{38}$ erg s$^{-1}$ was found to be coincident with the optical position of the outburst. 

LBVs represent a late phase of high mass ($\gtrsim20$ \Msun, Meynet \& Maeder 2005) stellar evolution during which the star undergoes significant of mass loss (up to 10$^{-4}$ \Msun yr$^{-1}$; see, e.g. Humphreys \& Davidson 1994) via stellar winds and/or strong shell ejections. LBV eruptions are typically detected as bright optical or infrared transients, and are frequently assigned an official SN designation only to be later recognized as ``impostors.'' X-ray emission observed during LBV eruptions is interpreted as shocks between ejected material from the erupting star and surrounding ISM (Smith 2008) or, in the case of the famous Galactic LBV \etaC, winds from a companion (likely an evolved O star or Wolf-Rayet star; Verner et al. 2005; Pittard \& Corcoran 2002). Colliding-wind binaries are the brightest class of stellar X-ray emitters that do not contain a compact companion (a neutron star or black hole). The brightest X-ray luminosities observed in these systems during outburst are on the order of $10^{35}$ erg s$^{-1}$ (Guerrero \& Chu 2008), three orders of magnitude lower than what was observed by {\it Swift} during the SN 2010da outburst.

Such high X-ray luminosities are commonly observed in high mass X-ray binaries (HMXBs). HMXBs consist of a compact object primary with a secondary massive star, typically either a Be or an OB supergiant. Be/X-ray binaries (BeXBs) typically contain a neutron star in a wide, moderately eccentric orbit around a Be companion. X-ray outbursts occur during periastron passage of the compact object. The majority of known HMXBs are BeXBs ($\sim$60\%, see Liu et al. 2006, and references therein), although the true fraction of BeXBs may be higher owing to the transient nature of the sources.

Supergiant X-ray binaries (SGXBs), on the other hand, consist of a compact object orbiting within the wind of a supergiant OB star. These systems show persistent X-ray emission (with X-ray luminosities of 10$^{35}$-10$^{36}$ erg s$^{-1}$) either from direct accretion of the stellar wind or from Roche-lobe overflow via an accretion disk. SGXBs are exceptionally rare: only a few dozen confirmed or suspected SGXBs are known in the Milky Way (Liu et al. 2006; Walter et al. 2006), and the low number of SGXB systems is naturally explained by the short lifetimes of the supergiant companion stars. 

In this Letter we report an X-ray detection of SN 2010da taken by the {\it Chandra} X-ray Observatory four months after the initial outburst and argue that SN 2010da is consistent with being an extragalactic SGXB. We describe our observations and data reduction procedures in \S2. The results of our imaging, hardness ratio analysis, spectral fitting, and time evolution since the outburst are presented in \S3. In \S4 we argue that SN 2010da is consistent with being an extragalactic SGXB, and a summary of our results is given in \S5.

\section{Observations and Data Reduction}
We have obtained a 63 ks observation of NGC 300 using the {\it Chandra} Advanced CCD Imaging Spectrometer (ACIS-I) on 2010 September 24. The X-ray observation was reduced using CIAO version 4.3 and CALDB version 4.4.2, using standard extraction procedures. We created exposure maps for the images using the CIAO script \texttt{merge\_all}\footnote{See http://cxc.harvard.edu/ciao/ahelp/merge\_all.html.}. Point sources were found using the CIAO task \texttt{wavdetect}\footnote{See http://cxc.harvard.edu/ciao/ahelp/wavdetect.html.}, and positions, positional errors, and 0.5-8 keV fluxes were measured for each point source, and spectra were extracted for all sources using \texttt{ACIS-Extract} (\texttt{AE}; Broos et al. 2010) version 2011-03-15. \texttt{AE} is a multi-purpose source extraction and characterization tool which determines source and background count rates, and calculates fluxes and source significance. SN 2010da was detected at $\sim$7$\sigma$ significance.
 
SN 2010da was detected at RA (J2000) = 00$^{\rm h}$55$^{\rm m}$04.$^{\rm s}$85 and Dec (J2000) = -37$^{\circ}$41\am43\as5, in excellent agreement (within 0\as2) with the position reported in SIMBAD\footnote{See http://simbad.u-strasbg.fr/.} four months after the outburst was first observed. The source is 3.8\am off-axis from the {\it Chandra} aim point. We detect a net 71 source counts in the 0.5-8 keV band. The background was estimated using an annular region, centered on the source position, extending from a radius of 25\asn to 30\asn away from the source. The 0.5-8 keV background count rate was found to be $\sim2\times10^{-6}$ ct s$^{-1}$ arcsec$^{-2}$. We assume a distance to NGC 300 of 2.0 Mpc (Dalcanton et al. 2009). All models include a column of neutral absorption fixed at the Galactic column $N_{\rm H,Gal}$ of $4.09\times10^{20}$ cm$^{-2}$ (Kalberla et al. 2005).

All spectral fitting was performed in \texttt{XSPEC} (Arnaud 1996) v.12.6.0q. We use $C$-statistics in lieu of traditional $\chi^2$ statistics due to the low number of source counts. Errors correspond to the 90\% confidence level. In addition to reporting the $C$-statistic per degrees of freedom (\Cdof) for each spectral model, we used the \texttt{XSPEC} command {\it goodness} to perform $10^4$ Monte Carlo realizations of the SN 2010da spectrum using our best-fit model. The command returns the percentage of simulated spectra that had a fit statistic less than that obtained from the fit to the real data -- a value of $\sim$50\% indicates the best-fit model is a good representation of the data. Percentages much smaller than 50\% indicate the data is over-parameterized by the model, and percentages much higher than 50\% indicate the model is a poor fit to  the data. We denote the Monte Carlo goodness-of-fit percentage as $MC$.

To derive an X-ray luminosity light curve, we supplement our measurements with archival X-ray data. The mid-outburst {\it Swift} XRT observations for SN 2010da are publicly available from the UK {\it Swift} Science Data Centre\footnote{See http://www.swift.ac.uk/user\_objects} (Evans et al. 2009), where data products (i.e., spectra and light curves) are automatically generated. SN 2010da was detected by XRT at RA (J2000) = 00$^{\rm h}$55$^{\rm m}$05.$^{\rm s}$02 and Dec (J2000) = -37$^{\circ}$41\am44\as5 with $\sim$270 counts. We use archival {\it XMM-Newton} images containing the SN 2010da progenitor within the field of view to estimate the 3$\sigma$ upper limit on the precursor X-ray luminosity using the \texttt{FLIX} web interface\footnote{See http://www.ledas.ac.uk/flix/flix.html.}. The upper limits determined by \texttt{FLIX} are found using the algorithm described in Carrera et al. (2007).

\section{Results}
\subsection{Imaging and Hardness Ratios}
Our {\it Chandra} observation was divided into three images consisting of different energy bands: ``soft'' (0.5-1 keV), ``medium'' (1-2 keV), and ``hard'' (2-8 keV). Each image was adaptively smoothed using the CIAO task \texttt{csmooth}\footnote{See http://cxc.harvard.edu/ciao/ahelp/csmooth.html}. Figure 1 shows a smoothed, RGB rendering of SN 2010da, with the source position and extraction radius ($\sim$2\asn) determined by \texttt{AE} superimposed. The extraction radius encompasses a hard point source, with extended soft emission.

We see evidence for a ring of soft X-ray emitting gas out to $\sim$2\asn ($\sim$20 pc) surrounding the central point source, with additional diffuse emission extended as far as $\sim$8\asn ($\sim$70 pc) north. The largest possible sphere of influence of the initial outburst would have a radius of only $\sim$0.1 pc, indicating the diffuse emission is the result of previous activity at earlier times in the system's evolutionary history. The size and shell-shaped structure of the diffuse emission is reminiscent of a supernova remnant (SNR). We therefore calculate the hardness ratio of the diffuse emission, using the definition from Prestwich et al. (2003) and the Plucinsky et al. (2008) catalog of SNRs in M~33:  $HR = (M-S)/(S+M+H)$, where $S$ is the soft 0.35-1.1 keV band,  $M$ is the medium 1.1-2.6 keV band, and $H$ is the hard 2.6-8.0 keV band. The diffuse emission has a $HR = -0.4$; using the  best-fit temperature of SNRs in M~33 ($kT\sim0.6$ keV), we estimate a 0.35-2 keV luminosity of $\sim6\times10^{35}$ erg s$^{-1}$.

The luminosity, hardness ratio, and size of the diffuse emission are all consistent with being a SNR; however, with only $\sim$10 net counts, we are unable to perform detailed spectral fitting to confirm this scenario. We examined public H$\alpha$ imaging of NGC~300\footnote{Images obtained via NED, http://nedwww.ipac.caltech.edu/}. There is a bright H$\alpha$ knot coincident with the tip of the shell, $\sim$8\asn north of SN 2010da. We find no obvious, bright H$\alpha$ emission coincident with the soft X-ray shell, making the source of the soft X-ray emission difficult to reliably classify.

We use the same energy bands described above to calculate the hardness ratio for the hard, central X-ray point source. We find $HR = 0.21\pm0.02$. Optical and infrared observations of the SN 2010da progenitor suggest the source was heavily absorbed, with $A_V>12$ mag of extinction (Berger \& Chornock 2010, ATEL \#2638). While significant intrinsic absorption would explain the  hard spectrum of SN 2010da, observations with the {\it Swift} Ultraviolet/Optical Telescope (UVOT) {\it during} the outburst showed the source was not heavily reddened (Brown 2010, ATEL \#2633), indicating that much of the dust enshrouding the progenitor was destroyed during the outburst. We discuss the level of intrinsic absorption in SN 2010da in more detail in the next section.

Given that SN 2010da appears to be a LBV centered on a SNR, we estimate the probability of a chance superposition using similarities in mass and star formation history between NGC~300 and M~33. We compared the locations of 37 LBVs (Massey et al. 2007) with 137 SNRs (Long et al. 2010) in M~33 and found no known M~33 LBV falls within a radius of less than 3\am of an M~33 SNR, suggesting that SN 2010da is not a chance superposition.

We also assess the probability of a chance superposition using NGC~300 itself. NGC~300 covers an area of $\sim$23 arcmin$^2$; if we assume there are $\sim$40 LBVs and $\sim$140 SNRs, the density of LBVs is 1.7 arcmin$^{-2}$ and the density of SNRs is 6 arcmin$^{-2}$. Therefore, a chance superposition within an area of 200 arcsec$^{2}$ has a probability of $\sim4\times10^{-4}$. This probability is a lower limit, since both types of objects would be of higher density in the star forming regions. Since no known LBV are associated with a SNR in M~33 inside an area of $\sim$0.5 deg$^2$, a reasonable estimate of the upper limit of the probability of a chance superposition is 0.2\%.

\subsection{Spectral Fitting}
X-ray emission originating from a colliding-wind binary would produce relatively soft thermal plasma spectrum. We therefore attempted to fit our 0.5-8 keV spectrum with a thermal plasma model ($kT$ = 10 keV), but found the fit unacceptable (\Cdof = 321/511 with $MC$ = 100\%). Allowing the temperature, abundances, or both to vary does not improve the goodness of the fit, and all two-temperature thermal plasma models over-parameterize the data. 

The best-fit spectral model from the {\it Swift} Data Centre pipeline is a power law with \PL = -0.03$^{+0.11}_{-0.08}$ and an unabsorbed 0.3-10 keV flux of 8.6$^{+1.6}_{-4.0}\times10^{-13}$ erg s$^{-1}$ cm$^{-2}$, implying an unabsorbed 0.3-10 keV luminosity of 4.5$^{+0.9}_{-2.1}\times10^{38}$ erg s$^{-1}$ at the time of the outburst. To directly compare our post-outburst observation with the {\it Swift} spectrum, we model the 0.5-8 keV spectrum with a power law and find a statistically acceptable fit with \PL = 0.15$\pm$0.42, with \Cdof = 276/511 and $MC$ = 58\%. We estimate an unabsorbed 0.3-10 keV flux of (4.5$\pm$2.5)$\times10^{-14}$ erg cm$^{-2}$ s$^{-1}$, corresponding to an unabsorbed luminosity of (2.4$\pm$1.3)$\times10^{37}$ erg s$^{-1}$. 

While a power law fit with \PL$\sim0$ is statistically acceptable, it is not physically motivated. When we include 12 mag of extinction (as found by Berger \& Chornock 2010, ATel \#2638), we find \PL= 2.18$^{+0.62}_{-0.64}$ with \Cdof = 297/510. However, $MC$ = 99.81\% and the fit residuals are poor, indicating an unacceptable fit. We find an absorbed disk black body model shows a similarly unacceptable fit to the data.

Observations of many Galactic SGXBs have shown the spectra to be consistent with a black body, with typical temperatures ranging from 1.6-1.9 keV (Sidoli et al. 2009). To test whether SN 2010da is consistent with being a wind-fed SGXB, we next use the \texttt{bbody} model in \texttt{XSPEC}. We find a best-fit temperature $kT = 1.79^{+0.66}_{-0.43}$ keV, with an unabsorbed 0.3-10 keV luminosity of 1.7$^{+0.2}_{-0.5}\times10^{37}$ erg s$^{-1}$ and \Cdof = 274/510. The model does not significantly over-parameterize the data, with $MC$ = 67.88\%. We find no evidence for absorption beyond the Galactic column, consistent with the majority of dust being destroyed during the outburst. The results of the spectral fitting are summarized in Table 1, and the best-fit spectral model and residuals are shown in Figure 2.

\subsection{Time Evolution}
We can further constrain the nature of the SN 2010da X-ray source using its time evolution. Multicolor optical photometry of SN 2010da in $B$, $V$, $R$, and $I$ filters was obtained during the outburst (Bond 2010, ATel \#2640) and nine days later (Prieto et al. 2010, ATel \#2660) on the SMARTS 1.3m telescope at Cerro Tololo using the ANDICAM camera. We calculate the $e$-folding time for each optical filter and the X-ray emission. We use the X-ray luminosity derived from the unabsorbed \PL$\sim0$ spectral model for both the {\it Swift} and {\it Chandra} observations to ensure the luminosity change is not due to a change in assumed spectral model between observations. We find $e$-folding times since the outburst for X-ray, $B$, $V$, $R$, and $I$ bands to be 41, 13, 18, 20, and 11 days, respectively. Compared to the decay rates of the optical magnitudes, the X-ray emission decay rate is significantly slower.

For comparison, we use optical monitoring from Fern\'{a}ndez-Laj\'{u}s et al. (2009) and X-ray {\it RXTE} monitoring from (Corcoran et al. 2010) to compute decay rates for \etaC. We find $e$-folding times for X-ray, $B$, $V$, $R$, and $I$ bands to be 71, 270, 220, 180, and 180 days, respectively -- nearly a factor of 2 longer in the X-rays than SN 2010da, and $\sim$10-20 times longer in the optical.

We have searched the {\it XMM-Newton} archive to constrain the level of X-ray emission of the SN 2010da progenitor. There are four archival observations (taken on 2000 Dec. 26, 2001 Jan. 1, 2005 May 22, and 2005 Nov. 25) containing the position of SN 2010da within the field of view. We use \texttt{FLIX} to retrieve the 3$\sigma$ upper limits on the 0.3-10 keV flux of the SN 2010da progenitor. During all four observations, the 3$\sigma$ upper limit of the unabsorbed 0.3-10 keV luminosity is estimated to be $\sim3-9\times10^{36}$ erg s$^{-1}$, two orders of magnitude lower than the observed outburst X-ray luminosity and $\sim$3 times lower than its current luminosity. These estimates are relatively insensitive to our choice of spectral model, and including heavy absorption as observed in the progenitor system changes our luminosity estimates by less than 20\%. Figure 3 shows all available X-ray data for SN 2010da and the progenitor system, including the 3$\sigma$ upper limits derived from the archival {\it XMM-Newton} images. 

Assuming our calculated X-ray $e$-folding time for SN 2010da, we expect SN 2010da to return to a ``quiescent'' state of 10$^{36}$ erg s$^{-1}$ in 8-9 months. This relatively slow rate of X-ray luminosity decay suggests that another source of X-ray emission keeps the X-ray luminosity high even as the massive star returns to a quiescent state. A wind-fed SGXB system naturally explains the time variability of SN 2010da -- the X-ray luminosity generated by the compact primary sharply increases as the secondary undergoes an outburst.

\section{SN 2010da: A Supergiant X-ray Binary?}
The best-fit black body spectrum of SN 2010da is consistent with the observed X-ray emission originating from known Galactic wind-fed SGXB. The X-ray luminosity of the progenitor system (less than $8\times10^{36}$ erg s$^{-1}$), as well as a sharp increase in X-ray emission coincident with an increased mass-loss rate of the secondary, further supports the SGXB scenario. 

We can roughly estimate the mass of the LBV assuming the SN 2010da outburst occurred when the massive star reached its Eddington limit, defined as $L_{\rm Edd}$ = 3.3$\times10^{4}$ ($M$/\Msun) $L_{\odot}$. The outburst optical luminosity is inferred from optical magnitudes to be $\sim10^6$ $L_{\odot}$, and the SED of SN 2010da nine days post-outburst can be modeled as a black body with a luminosity of $\sim$5.5$\times10^5$ $L_{\odot}$ (Prieto et al. 2010, ATel \#2660). We therefore estimate the mass of the LBV to be $\sim20-30$ \Msun. Assuming that the star has a quiescent absolute $V$-band magnitude of -8.6 and $B-V=0.4$ (roughly the values measured $\sim$1-2 weeks after outburst), we use the isochrones of Girardi et al. (2002) to estimate the mass of the LBV to be roughly 24 \Msun, consistent with the Eddington limit estimates. We note that this estimate does not use the true quiescent magnitudes of the system, and the uncertainties in the current level of dust extinction may affect our mass estimates. Better constraints on the quiescent optical luminosity are needed to refine our mass estimate.

Optical and IR observations of the SN 2010da progenitor showed a clear mid-infrared (MIR) excess, with an intrinsic absorption of $A_V>12$ mag (\nH$\gtrsim4\times10^{22}$ cm$^{-2}$) required to explain the non-detection of the source at optical wavelengths. The SED of the progenitor system was well-described by a black body, with $T_d=890$ K and $R_d=22 R_{\star}$ (Prieto et al. 2010, ATel \#2660). Rahoui et al. (2008) combined optical, near-infrared (NIR), and MIR photometry for 12 well-studied Galactic SGXBs to perform spectral energy distribution (SED) fitting. These sources exhibited heavy intrinsic absorption and MIR excess and were well-described by a simple stellar black body model with a cocoon of dust and/or cold gas enshrouding the system (three sources required an additional warm dust component); typical dust temperatures and radii were found to be $T_d\sim800-1100$ K and $R_d\sim10R_{\star}$. Observations of the SN 2010da progenitor are therefore consistent with observations of Galactic SGXBs.

\section{Summary}
We present the results of a 63 ks {\it Chandra} observation of the LBV SN 2010da, four months after it was initially detected on 24 May 2010. We detect an X-ray source at $\sim$7$\sigma$ confidence coincident with the optical location of the SN 2010da outburst. Our imaging analysis has revealed a hard central point source, surrounded by soft diffuse emission extending as far as $\sim$8\asn north of the central source. The hardness ratio ($HR=-0.4$), luminosity ($\sim6\times10^{35}$ erg s$^{-1}$), and size ($\sim20$ pc) of the diffuse emission are all consistent with being a typical SNR in M~33 (Plucinsky et al. 2008). However, with only $\sim10$ net counts, detailed spectral modeling to confirm a SNR origin of the X-ray emission is impossible.

The central point source exhibits significantly harder X-ray emission, with $HR=0.21$. While optical and infrared observations of the SN 2010da progenitor was heavily extincted, with $A_V>12$ mag, optical and ultraviolet observations during the outburst indicate that much of the absorbing dust was destroyed during the outburst. Our preferred spectral model is a black body ($kT = 1.79^{+0.66}_{-0.43}$) with no intrinsic absorption beyond the Galactic column. We estimate the 0.3-10 keV luminosity to be 1.7$^{+0.2}_{-0.5}\times10^{37}$ erg s$^{-1}$, a factor of $\sim$25 decrease since the initial outburst four months previously. We use archival {\it XMM-Newton} images to constrain the 3$\sigma$ upper limit X-ray luminosity of the progenitor to be $3-9\times10^{36}$ erg s$^{-1}$. 

A deep follow-up X-ray observation, sensitive to a limiting luminosity of a few 10$^{35}$ erg s$^{-1}$ (like the observation we have obtained here) would be capable of verifying the quiescent X-ray luminosity of this system, confirm the SGXB origin of the X-ray emission, and may provide further evidence for a supernova remnant origin of the diffuse emission. Time-resolved optical spectroscopy of the massive star may confirm the presence of a compact object primary and provide constraints on the masses and orbital parameters of this system, and optical and infrared photometric monitoring could constrain the rate of dust regeneration in the system.

\acknowledgements
This work made use of data supplied by the UK {\it Swift} Science Data Centre at the University of Leicester and the NASA/IPAC Extragalactic Database (NED) which is operated by the Jet Propulsion Laboratory, California Institute of Technology, under contract with the National Aeronautics and Space Administration.  B. B. and B. F. W. acknowledge support from {\it Chandra} grant  GO1-12118X. T. J. G. and P.P.P acknowledge support of NASA Contract NAS8-03060. The authors would like to thank M. Eracleous and E. Skillman for helpful comments on early drafts of this manuscript.

%%%%%TABLES
\begin{table}[!ht]\footnotesize
\caption{Summary of 0.5-8 keV Spectral Fitting}
\centering
\begin{tabular}{m{2.5cm} lll}
\hline\hline
Model & Parameter & Best-Fit Value & Units \\
\hline
\multirow{4}{2.5cm}{{\it Swift} power law, during outburst} & \PL  & -0.03$^{+0.11}_{-0.08}$ & \\
   & unabs. flux  & 8.6$^{+1.6}_{-4.0}$ & $10^{-13}$ erg s$^{-1}$ cm$^{-2}$ \\
   & unabs. luminosity  & 4.5$^{+0.9}_{-2.1}$ & $10^{38}$ erg s$^{-1}$    \\
   & \Cdof                    & 255/237 & \\
\hline
\multirow{6}{2.5cm}{{\it Chandra} power law, post-outburst} & \nH & $4.09\times10^{20}$ (fixed) & cm$^{-2}$ \\ 
 &  \PL & 0.15$\pm$0.42 & \\
 & unabs. flux & 4.5$\pm$2.5 & $10^{-14}$ erg s$^{-1}$ cm$^{-2}$ \\
   & unabs. luminosity   & 2.4$\pm$1.3 & $10^{37}$ erg s$^{-1}$   \\
   & \Cdof                                             & 276/511 & \\
   & $MC$                                               & 58.14\% & \\
\hline
\multirow{6}{2.5cm}{{\it Chandra} black body, post-outburst} & \nH & $4.09\times10^{20}$ (fixed) & cm$^{-2}$ \\ 
  & $kT$ & 1.79$^{+0.66}_{-0.43}$ & keV \\ 
 & unabs. flux & 3.2$^{+0.4}_{-0.9}$ & $10^{-14}$ erg s$^{-1}$ cm$^{-2}$ \\
   & unabs. luminosity & 1.7$^{+0.2}_{-0.5}$ & $10^{37}$ erg s$^{-1}$   \\
                        & \Cdof                                             & 274/510 & \\
                       & $MC$                                               & 67.88\% & \\
\hline\hline
\end{tabular}
\end{table}

%\begin{table}[!ht]\footnotesize
%\caption{SN 2010da Decay Times Compared to \etaC}
%\centering
%\begin{tabular}{cccccc}
%\hline\hline
%           & \multicolumn{2}{c}{SN 2010da} && \multicolumn{2}{c}{\etaC} \\ \cline{2-3} \cline{5-6}
%Band & Peak Luminosity & $e$-Folding Time && Peak Luminosity & $e$-Folding Time \\
%          &  or Absolute Magnitude  &   (days) && or Absolute Magnitude     & (days) \\
%\hline
%X-ray &   4.5$\times10^{38}$ erg s$^{-1}$  & 41 & & $\sim10^{35}$ erg s$^{-1}$ & 71 \\
%$B$ &   -9.9 mag & 13 && -6.3 mag  & 270 \\
%$V$ & -10.2 mag & 18 && -8.0 mag  & 220 \\
%$R$ & -10.6 mag & 20 && -7.1 mag  & 180 \\
%$I$   & -10.8 mag & 11 && -8.6 mag  & 180 \\
%\hline\hline
%\end{tabular}
%\end{table}

%%%%%FIGURES
\begin{figure}[!ht]
\centering
\begin{tabular}{c}
\includegraphics[width=0.6\linewidth]{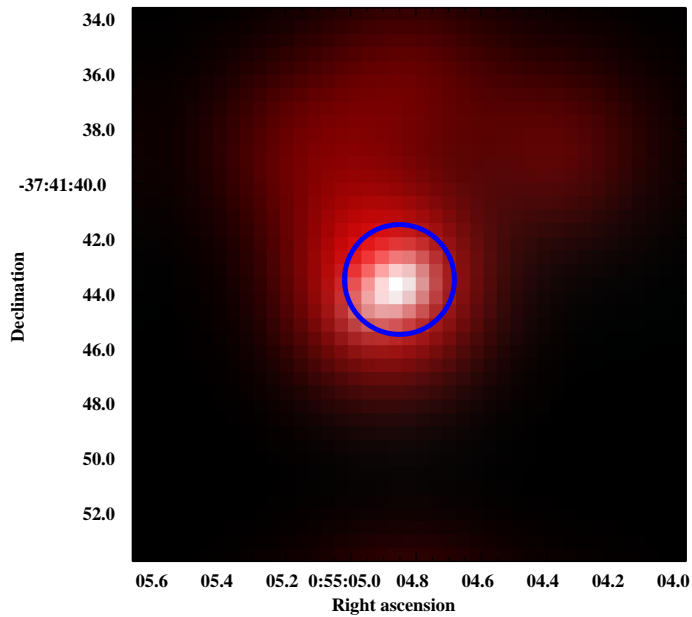} 
\end{tabular}
\caption{An adaptively smoothed image of our {\it Chandra} observation of SN 2010da. This rendering emphasizes the hard point source and diffuse soft emission. The source position from \texttt{wavdetect} and extraction radius determined by \texttt{ACIS-Extract} is shown by the thick blue circle. Red = 0.5-1 keV, green = 1-2 keV, and blue = 2-8 keV.}
\end{figure}

\begin{figure}[!ht]
\centering
\begin{tabular}{c}
\includegraphics[width=0.5\linewidth,angle=-90]{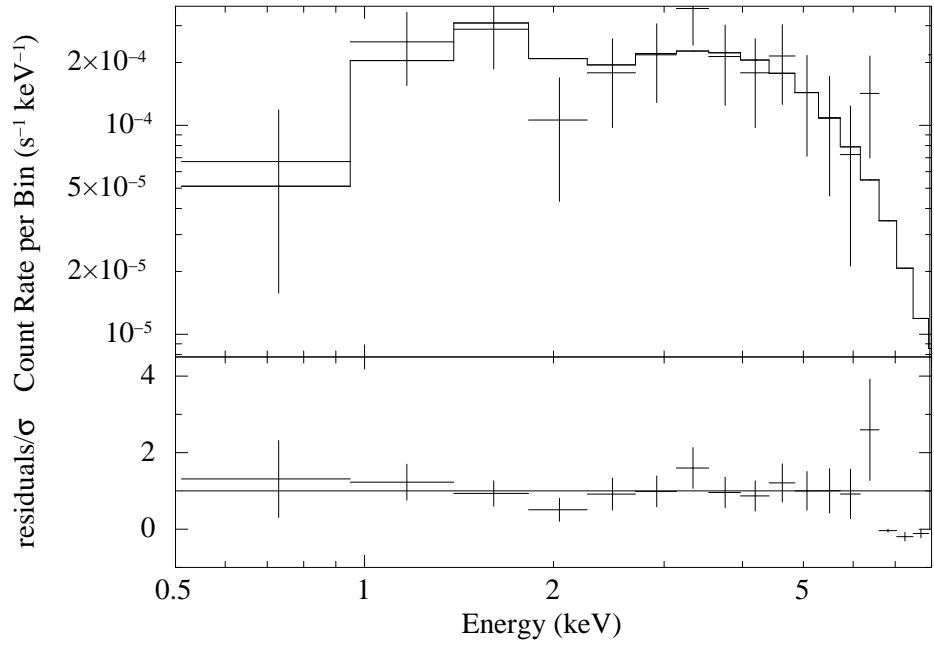} \\
\end{tabular}
\caption{The {\it Chandra} 0.5-8.0 keV spectrum of SN 2010da. Our best-fit spectral model, a black body emitter ($kT=1.79^{+0.68}_{-0.43}$ keV) with no evidence for additional absorption beyond the Galactic column, is superimposed.}
\end{figure}

%\begin{figure}[!ht]
%\centering
%\begin{tabular}{c}
%\includegraphics[width=0.75\linewidth,clip=true,trim=1cm 12cm 1cm 1cm]{decay.pdf} \\
%\caption{The luminosity decay of SN 2010da over a one month period. The X-ray luminosity has been arbitrarily scaled to be shown on the same axes as the optical magnitudes.}
%\end{tabular}
%\end{figure}

\begin{figure}[!ht]
\centering
\begin{tabular}{c}
\includegraphics[width=0.7\linewidth]{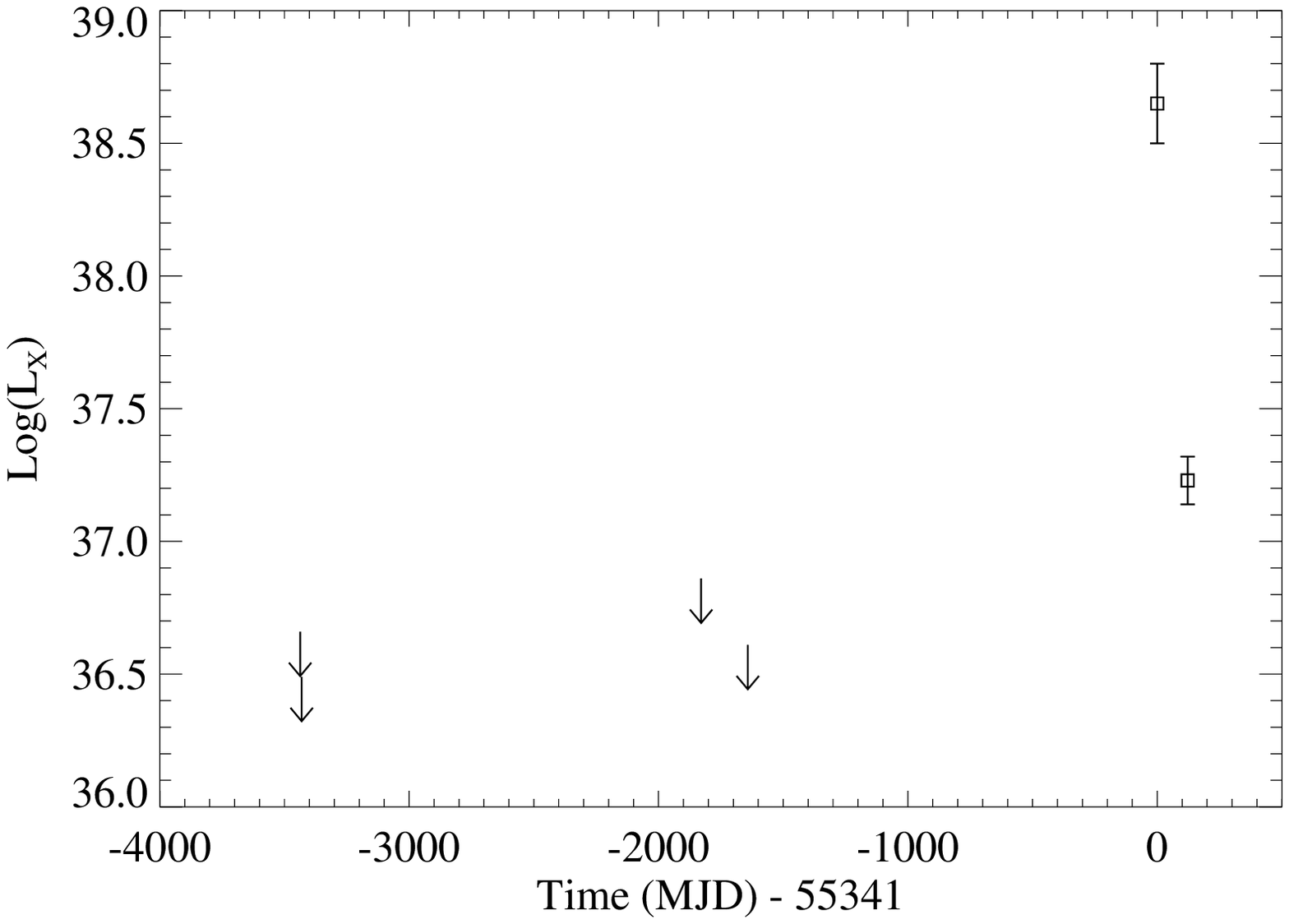} \\
\caption{The long-term X-ray luminosity of SN 2010da. Downward arrows indicate 3$\sigma$ upper limits derived from archival {\it XMM-Newton} images. The time axis is relative to the date of the outburst.}
\end{tabular}
\end{figure}


\begin{references}
\reference{Be10} Berger, E. \& Chornock, R. 2010, ATel \# 2638
\reference{Bo10} Bond, H. E. 2010, ATel \#2640
\reference{Br10} Broos, P. S., Townsley, L. K., Feigelson, E. D., Getman, K. V., Bauer, F. E., \& Garmire, G. P. 2010, ApJ, 714, 1582
\reference{Br10} Brown, P. J. 2010, ATel \#2633
\reference{Ca07} Carrera, F. J. et al. 2007, A\&A, 469, 27
\reference{Co98} Corcoran, M. F. et al. 1998, ApJ, 494, 381
\reference{Co00} Corcoran, M. F., Fredericks, A. C., Petre, R., Swank, J. H., \& Drake, S. A. 2000, ApJ, 545, 420
\reference{Co01} Corcoran, M. F. et al. 2001, ApJ, 562, 1031  
\reference{Co05} Corcoran, M. F. 2005, AJ, 129, 2018
\reference{Co10} Corcoran, M. F., Hamaguchi, K., Pittard, J. M., Russell, C. M. P., Owocki, S. P.; Parkin, E. R., Okazaki, A.	2010, ApJ, 725, 1528
\reference{Da09} Dalcanton, J. J. et al. 2009, ApJS, 183, 67
\reference{ER10} Elias-Rosa, N., Mauerhan, J. C. \& Van Dyk, S. D. 2010, ATel \#2636
\reference{FL09} Fern\'{a}ndez-Laj\'{u}s, E. et al. 2009, A\&A, 493, 1093
\reference{Ga09} Gal-Yam, A. \& Leonard, D. C. 2009, Nature, 458, 865
\reference{Gi02} Girardi, L., Bertelli, G., Bressan, A., Chiosi, C., Groenewegen, M. A. T., Marigo, P., Salasnich, B., \& Weiss, A. 2002, A\&A, 391, 195
\reference{Gu08} Guerrero, M. A. \& Chu, Y.-H. 2008, ApJS, 177, 216
\reference{Ka05} Kalberla, P. M. W., Burton, W. B., Hartmann, D., Arnal, E. M., Bajaja, E., Morras, R., 
\& P\"{o}ppel, W. G. L. 2005, A\&A, 440, 775
\reference{Kh10} Khan, R., Stanek, K. Z., Kochanek, C. S., Thompson, T. A. \& Prieto, J. L. 2010, ATel \#2632
\reference{Hu94} Humphreys, R. M. \& Davidson, K. 1994, PASP, 106, 1025
\reference{Im10} Immler, S., Brown, P. J. \& Russell, B. R. 2010, ATEL \#2639
%\reference{Ma92} Maeder, A. 1992, Instabilities in Evolved Super- and Hypergiants, Royal Netherlands Academy of Arts and Sciences, p. 138
\reference{Li00} Liu, Q. Z., van Paradijs, J., \& van Heuvel, E. P. J. 2000, A\&A,S, 147, 25
\reference{Lo10} Long, K. S. et al. 2010, ApJS, 187, 495
\reference{Ma07} Massey, P., McNeill, R. T., Olsen, K. A. G., Hodge, P. W., Blaha, C., Jacoby, G. H., Smith, R. C., \& Strong, S. B. 2007, AJ, 134, 2474
\reference{Me05} Meynet, G. \& Maeder, A. 2005, A\&A, 581, 598
\reference{Mi10} Miller, A. A. et al. 2010, MNRAS, 404, 305
\reference{Pa09} Parkin, E. R., Pittard, J. M., Corcoran, M. F., Hamaguchi, K., \& Stevens, I. R. 2009, MNRAS, 394, 1758	
\reference{Pi02} Pittard, J. M. \& Corcoran, M. F. 2002, A\&A, 383, 636
\reference{Pl08} Plucinsky, P. P. et al. 2008, ApJS, 174, 366
\reference{Pr03} Prestwich, A. H., Irwin, J. A., Kilgard, R. E., Krauss, M. I., Zezas, A., Primini, F., Kaaret, P., Boroson, B. 2003, ApJ, 595, 719
\reference{Pr10} Prieto, J. L., Bond, H. E., Kochanek, C. S., Khan, R., Stanek, K. Z., \& Thompson, 
T. A. 2010, ATel \#2660
\reference{Ra08} Rahoui, F., Chaty, S., Lagage, P.-O., \& Pantin, E. 2008, A\&A, 484, 801
\reference{Sc98} Schlegel, D. J., Finkbeiner, D. P., \& Davis, M. 1998, ApJ, 500, 525
\reference{Sc09} Schinzel, F. K.,  Taylor, G. B.,  Stockdale, C. J., Granot, J., \& Ramirez-Ruiz, E. 2009, ApJ, 691, 1380
\reference{Se09} Sekiguchi, A., Tsujimoto, M., Kitamoto, S., Ishida, M., Hamaguchi, K., Mori, H., Tsuboi, Y.  2009, PASJ, 61, 629
\reference{Si09} Sidoli, L. et al. 2009, MNRAS, 397, 1528
\reference{Sm07} Smith, N. \& McCray, R. 2007, ApJ, 671, L17
\reference{Sm07} Smith, N. et al 2007, ApJ, 666, 1116
\reference{Sm08} Smith, N. 2008, Nature, 455, 201
\reference{Sm09} Smith, N., Chornock, R., Li, W., Ganeshalingam, M., Silverman, J. M., Foley, R. J., Filippenko, A. V., \& Barth, A. J. 2009, ApJ, 686, 467
\reference{Sm10} Smith, N. et al. 2010, AJ, 139, 1451
\reference{Ti05} Tikhonov, N. A., Galazutdinova, O. A., \& Drozdovsky, I. O. 2005, A\&A, 431, 127
\reference{Ve05} Verner, E., Bruhweiler, F., \& Gull, T. 2005, ApJ, 624, 973
\reference{Wa06} Walter, R. et al. 2006, A\&A, 453, 133
\reference{We01} Weis, K., Duschl, W. J. \& Bomans, D. J. 2001, A\&A, 367, 566
\end{references}
\end{document}